# Deep-learning-driven Reliable Single-pixel Imaging with Uncertainty Approximation

**Ruibo Shang, Mikaela A. O'Brien, and Geoffrey P. Luke[*]**

*Thayer School of Engineering, Dartmouth College, 14 Engineering Dr., Hanover, NH 03755, USA*

*[*]geoffrey.p.luke@dartmouth.edu*

## Abstract

Single-pixel imaging (SPI) has the advantages of high-speed acquisition over a broad wavelength range and system compactness, which are difficult to achieve by conventional imaging sensors. However, a common challenge is low image quality arising from undersampling. Deep learning (DL) is an emerging and powerful tool in computational imaging for many applications and researchers have applied DL in SPI to achieve higher image quality than conventional reconstruction approaches. One outstanding challenge, however, is that the accuracy of DL predictions in SPI cannot be assessed in practical applications where the ground truths are unknown. Here, we propose the use of the Bayesian convolutional neural network (BCNN) to approximate the uncertainty (coming from finite training data and network model) of the DL predictions in SPI. Each pixel in the predicted result from BCNN represents the parameter of a probability distribution rather than the image intensity value. Then, the uncertainty can be approximated with BCNN by minimizing a negative log-likelihood loss function in the training stage and Monte Carlo dropout in the prediction stage. The results show that the BCNN can reliably approximate the uncertainty of the DL predictions in SPI with varying compression ratios and noise levels. The predicted uncertainty from BCNN in SPI reveals that most of the reconstruction errors in deep-learning-based SPI come from the edges of the image features. The results show that the proposed BCNN can provide a reliable tool to approximate the uncertainty of DL predictions in SPI and can be widely used in many applications of SPI.

# 1. Introduction

Single-pixel imaging (SPI) [1] is a novel imaging technique which uses a single-element photodetector to record the image information instead of using pixel array image sensors such as charge-coupled device (CCD) and complementary metal-oxide-semiconductor (CMOS) cameras. The object is sequentially illuminated by a set of special designed patterns and the total intensity light for each pattern illumination is collected as a single-pixel value by the photodetector. Finally, the computational algorithms are applied to reconstruct the object with the sequential intensity collections and the illumination patterns. SPI has many advantages including high speed, broad bandwidth and compact imaging [2]. It also has many applications including remote sensing [3], holography [4, 5], optical encryption [6, 7] and tomography [8].

One of the most common reconstruction methods in SPI is sparsity-based optimization which seeks to reconstruct images from incomplete measurements [9, 10] by incorporating the knowledge that most natural images are sparse (i.e., only a few nonzero values exist) when the image is transformed into a specific domain. However, the primary drawback to this approach is that it is time consuming because of its iterative nature. Depending on the scale of the model and scope of the problem, an image reconstruction task can take minutes to even hours to compute. Therefore, real-time imaging is infeasible for applications that require pipelined data acquisition and image reconstruction. Besides, the optimal algorithm-specific parameters (i.e., the regularization parameter) in the sparsity-based optimization framework generally need to be heuristically determined.

Deep learning (DL) [11, 12] is an emerging and powerful computational imaging approach dramatically improving the state-of-the-art in image reconstruction compared with conventional reconstruction algorithms [13-19]. It relies on large amounts of training data to automatically learn tasks by finding the optimal weights in each layer of a neural network [12]. This is in contrast to iteratively optimizing the image with a specific model in sparsity-based optimization approaches. Therefore, DL is a promising alternative to augment or replace iterative algorithms used in sparsity-based optimization. Researchers have applied DL approaches in SPI to improve the quality of the reconstructed images compared with conventional approaches [20-25]. For instance, a DL approach was proposed to predict the image in SPI with an initial guess of the image from conventional approaches as the input to the DL network to further improve the image quality at high compression ratios and noise levels [21]. End-to-end DL approaches [22, 24] were proposed to predict the image in SPI directly from the raw measurement data without the knowledge of the imaging model and therefore no pre-processing of the raw measurement data is needed.

Generally, the accuracy of the DL predictions in SPI can be quantified by comparing with the ground-truth images [22] (i.e., calculating mean absolute error (MAE), root mean squared error (RMSE) and structural similarity index (SSIM) [26]). However, one outstanding challenge is that the ground truths are usually unknown during the prediction stage in many practical applications. Therefore, the accuracy of the DL prediction of a specific image cannot be estimated.

Bayesian convolutional neural networks (BCNN) has been proved to be an effective approach to approximate the uncertainty with many applications including image segmentation [27], phase imaging [16] and image classification [28]. BCNN work on the principle that each pixel in the output image represents the parameter of a probability distribution (e.g., Laplacian or Gaussian distribution), rather than a single intensity value. Then, the uncertainties can be quantified by Monte Carlo dropout [29] and the Deep Ensembles [30] etc.

In this paper, we propose to use a BCNN in SPI to simultaneously predict the image and the pixel-wise uncertainty to approximate the accuracy of the predicted image. In Section 2, we model the uncertainties (data uncertainty, model uncertainty and the overall uncertainty) with the Laplacian-distributed, Gaussian-distributed and Bernoulli-distributed likelihood functions. Respectively (*Section 2.1*), and propose the networks structures for BCNN in SPI (*Section 2.2*). Besides, the data simulation and pre-processing is described in *Section 2.3* and the experimental data acquisition and pre-processing is described in *Section 2.4*. In Section 3, we show the BCNN predictions of the image and uncertainty in simulated SPI with analysis in details. Overall, these results show that uncertainty approximation can be used to reliably interpret the result of a compressed computational imaging problem.

## 2. Methods

### *2.1 Bayesian networks for uncertainty approximation*

As opposed to conventional convolutional neural networks where the weights are deterministic after training, BCNNs use distributions over the network parameters to replace the deterministic weights in the network. This probabilistic property of BCNN come from the stochastic (random) processes in the network such as dropout [31], weight initialization [32] etc. Suppose the training dataset is denoted as $(X, Y) = \{x_n, y_n\}_{n=1}^{N}$ with $X$ and $Y$ representing the network inputs and ground-truth images, respectively. N is the total number of images in the training dataset. To approximate the variability of the prediction y given a specific input $x_{test,t}$ in the testing dataset $(X_{test}, Y_{test}) = \{x_{test,t}, y_{test,t}\}_{t=1}^{T}$ (T is the total number of images in the testing dataset), we use the predictive distribution $p(y|x_{test,t}, X, Y)$ over all possible learned weights (with marginalization) [29]:

$$p(y|x_{test,t}, X, Y) = \int p(y|x_{test,t}, W) p(W|X, Y) dW \tag{1}$$

where $p(y|x_{test,t}, W)$ denotes the predictive distribution that includes all possible output predictions given the learned weights $W$ and the input $x_{test,t}$ from the testing dataset. It can be understood as data uncertainty [16]. $p(W|X, Y)$ denotes all possible learned weights given the training dataset, which can be understood as model uncertainty [16].

To model the data uncertainty, we need to define the probability distribution of the BCNN outputs with a specific likelihood function. In this paper, we choose the multivariate Laplacian-distributed, Gaussian-distributed and Bernoulli-distributed likelihood functions to model the data uncertainty.

(a) We define the multivariate Laplacian-distributed likelihood function as:

$$p_{Laplacian}(y|x, W) = \prod_{m=1}^{M} p(y^m|x, W) \tag{2}$$

$$p_{Laplacian}(y^m|x, W) = \frac{1}{2\sigma^m} \exp\left(-\frac{|y^m - \mu^m|}{\sigma^m}\right) \tag{3}$$

where m denotes the mth pixel in the BCNN output image, M denotes the total number of pixels in the BCNN output image, and $\mu^m$ and $\sigma^m$ denote the mean and standard deviation of the mth pixel in the BCNN output image, respectively.

By taking logarithm and negative operations on Eq. (2), the loss function $L_{Laplacian}(W|x_n, y_n)$ for the Laplacian-distributed likelihood function given the training data pair $(x_n, y_n)$ is:

$$L_{Laplacian}(W|x_n, y_n) = \frac{1}{M} \sum_{m=1}^{M} \left[\frac{|y_n^m - \mu_n^m|}{\sigma_n^m} + \log(2\sigma_n^m)\right] \tag{4}$$

(b) For multivariate Gaussian-distributed likelihood function, we define:

$$p_{Gaussian}(y|x, W) = \prod_{m=1}^{M} p(y^m|x, W) \tag{5}$$

$$p_{Gaussian}(y^m|x, W) = \frac{1}{\sqrt{2\pi}\sigma^m} \exp\left[-\frac{(y^m - \mu^m)^2}{2(\sigma^m)^2}\right] \tag{6}$$

where the denotations are the same as those in Eq. (2) and (3).

By taking logarithm and negative operations on Eq. (5), the loss function $L_{Gaussian}(W|x_n, y_n)$ for the Gaussian-distributed likelihood function given the training data pair $(x_n, y_n)$ is:

$$L_{Gaussian}(W|x_n, y_n) = \frac{1}{M} \sum_{m=1}^{M} \left[ \frac{(y_n^m - \mu_n^m)^2}{2(\sigma_n^m)^2} + \log(\sqrt{2\pi}\sigma_n^m) \right] \qquad (7)$$

(c) For Bernoulli-distributed likelihood function, we define:

$$p_{Bernoulli}(y|x, W) = \prod_{m=1}^{M} p(y^m|x, W) \qquad (8)$$

$$p_{Bernoulli}(y^m = 1|x, W) = \mu^m \qquad (9)$$

$$p_{Bernoulli}(y^m|x, W) = (\mu^m)^{y^m}(1-\mu^m)^{1-y^m} \qquad (10)$$

where the denotations are the same as those in Eq. (2) and (3).

By taking logarithm and negative operations on Eq. (8), the loss function $L_{Bernoulli}(W|x_n, y_n)$ for the Bernoulli-distributed likelihood function given the training data pair $(x_n, y_n)$ is:

$$L_{Bernoulli}(W|x_n, y_n) = \sum_{m=1}^{M} [y_n^m - 1)\log(1 - \mu_n^m) - y_n^m \log(\mu_n^m)] \qquad (11)$$

We would like to learn the weights to maximize Eqs. (2), (5) and (8) in the training dataset, which is equivalent to minimizing the loss functions defined in Eqs. (4), (7) and (11). There are two channels ($\mu$ and $\sigma$) in the BCNN output for Laplacian-distributed and Gaussian-distributed likelihood functions while there is only one channel ($\mu$) in the BCNN output for the Bernoulli-distributed likelihood function.

To measure the model uncertainty, we use the dropout network [29]. A specifically defined distribution $q(W)$ [29] is learned to approximate $p(W|X, Y)$ (minimizing the Kullback-Leibler divergence between $q(W)$ and $p(W|X, Y)$) by applying a dropout layer before every layer that has learnable weights. During the prediction process, the model uncertainty is approximated by Monte Carlo dropout [29]. With Monte Carlo integration, the predictive distribution $p(y|x_{test,t}, X, Y)$ in Eq. (1) can be approximated as:

$$p(y|x_{test,t}, X, Y) \approx \int p(y|x_{test,t}, W) q(W) dW \approx \frac{1}{K} \sum_{k=1}^{K} p(y|x_{test,t}, W^k) \qquad (12)$$

where K is the total number of times of dropout activations during the prediction process.

Finally, the predicted image can be represented by the predicted mean $\hat{\mu}_{test,t}^m$ of the mth pixel for the testing data $x_{test,t}$ (for Laplacian-distributed, Gaussian-distributed and Bernoulli-distributed likelihood functions) is:

$$\hat{\mu}_{test,t}^{m} = \mathbb{E}[y^m|x_{test,t}, X, Y] \approx \frac{1}{K}\sum_{k=1}^{K}\mathbb{E}[y^m|x_{test,t}, W^k] \approx \frac{1}{K}\sum_{k=1}^{K}\hat{\mu}_{test,t}^{m,k} \quad (13)$$

where $\mathbb{E}$ denotes the expectation and $\hat{\mu}_{test,t}^{m,k}$ denotes the predicted $\mu$ of the mth pixel and kth dropout activation for the testing data $x_{test,t}$.

The predicted uncertainty $\hat{\sigma}_{test,t}^{m}$ of the mth pixel for the testing data $x_{test,t}$ for Laplacian-distributed likelihood function is:

$$\begin{aligned}\hat{\sigma}_{test,t(Laplacian)}^{m} &= \sqrt{Var(y^m|x_{test,t}, X, Y)} \\ &= \sqrt{\mathbb{E}[Var(y^m|x_{test,t}, W, X, Y)] + Var(\mathbb{E}[y^m|x_{test,t}, W, X, Y])} \\ &= \sqrt{\mathbb{E}[Var(y^m|x_{test,t}, W)] + Var(\mathbb{E}[y^m|x_{test,t}, W])} \\ &\approx \sqrt{\frac{1}{K}\sum_{k=1}^{K}2(\hat{\sigma}_{test,t}^{m,k})^2 + \frac{1}{K}\sum_{k=1}^{K}(\hat{\mu}_{test,t}^{m,k} - \hat{\mu}_{test,t}^{m})^2} = \sqrt{(\hat{\sigma}_{test,t}^{m(D)})^2 + (\hat{\sigma}_{test,t}^{m(M)})^2}\end{aligned} \quad (14)$$

where $Var$ denotes pixel-wise variance, $\hat{\sigma}_{test,t}^{m,k}$ denotes the predicted standard deviation of the mth pixel and kth dropout activation for the testing data $x_{test,t}$. $\hat{\sigma}_{test,t}^{m(D)} = \sqrt{\frac{1}{K}\sum_{k=1}^{K}2(\hat{\sigma}_{test,t}^{m,k})^2}$ denotes the data uncertainty and $\hat{\sigma}_{test,t}^{m(M)} = \sqrt{\frac{1}{K}\sum_{k=1}^{K}(\hat{\mu}_{test,t}^{m,k} - \hat{\mu}_{test,t}^{m})^2}$ denotes the model uncertainty.

For Gaussian-distributed likelihood function:

$$\begin{aligned}\hat{\sigma}_{test,t(Gaussian)}^{m} &\approx \sqrt{\frac{1}{K}\sum_{k=1}^{K}(\hat{\sigma}_{test,t}^{m,k})^2 + \frac{1}{K}\sum_{k=1}^{K}(\hat{\mu}_{test,t}^{m,k} - \hat{\mu}_{test,t}^{m})^2} \\ &= \sqrt{(\hat{\sigma}_{test,t}^{m(D)})^2 + (\hat{\sigma}_{test,t}^{m(M)})^2}\end{aligned} \quad (15)$$

where the denotations are the same as those in Eq. (14) and the derivation of Eq. (15) is similar to that of Eq. (14).

For Bernoulli-distributed likelihood function:

$$\begin{aligned}\hat{\sigma}_{test,t(Bernoulli)}^{m} &\approx \sqrt{\frac{1}{K}\sum_{k=1}^{K}[\hat{\mu}_{test,t}^{m,k}(1 - \hat{\mu}_{test,t}^{m,k})] + \frac{1}{K}\sum_{k=1}^{K}(\hat{\mu}_{test,t}^{m,k} - \hat{\mu}_{test,t}^{m})^2} \\ &= \sqrt{(\hat{\sigma}_{test,t}^{m(D)})^2 + (\hat{\sigma}_{test,t}^{m(M)})^2}\end{aligned} \quad (16)$$

where the denotations are the same as those in Eq. (14) and the derivation of Eq. (16) is similar to that of Eq. (14).

We can find from Eq. (14-16) that the data uncertainty ($\hat{\sigma}_{test,t}^{m(D)}$) is approximated by the mean of the predicted variance and the model uncertainty ($\hat{\sigma}_{test,t}^{m(M)}$) is approximated by the variance of the predicted mean.

## 2.2. BCNN structures

The BCNN structures are shown in Fig. 1. They follow the U-Net architecture [33], which utilizes an encoder-decoder structure with skip connections to preserve wide-frequency features. This architecture was chosen because of its success in solving image-to-image problems. Dropout layers with a dropout rate of 0.1 were included before each convolution layer of the U-Net. $L_2$ kernel regularizer and bias regularizer with the regularization factor of $1 \times 10^{-6}$ were included in each convolution layer. The network structure in Fig. 1 is used for Bernoulli-distributed likelihood function and the network structure used for both Laplacian-distributed and Gaussian-distributed likelihood functions is the same as the one in Fig. 1 except that there are two output channels. The loss functions in Eq. (4), (7) and (11) were used in BCNN for Laplacian-distributed, Gaussian-distributed and Bernoulli-distributed likelihood function, respectively. The BCNN was trained on a NVIDIA Quadro M4000 GPU with an 8GB of memory.

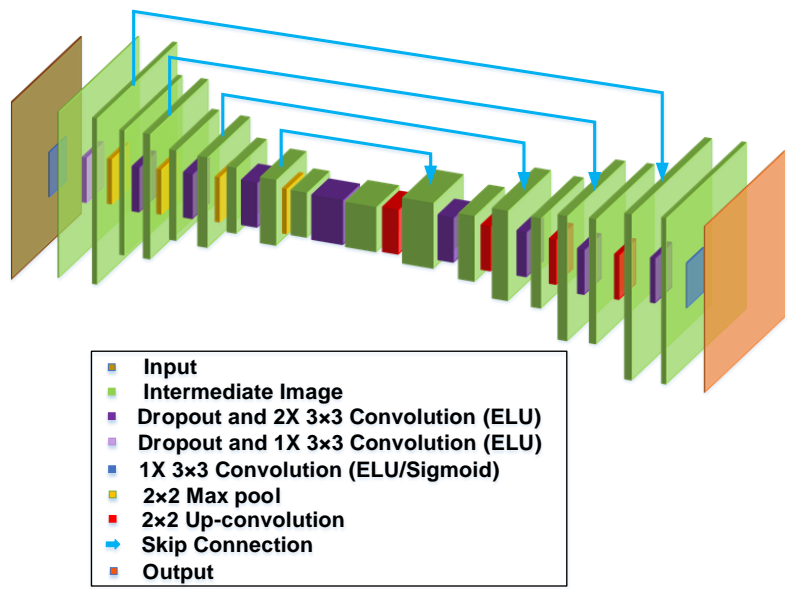

Fig. 1 The BCNN structure.

## 2.3. Data simulation and pre-processing

Russian-Doll (RD) Hadamard [34] patterns are used as the sampling patterns in the simulated SPI. In RD Hadamard patterns, the measurement order of the Hadamard basis is reordered and optimized according to their significance for general scenes, such that at discretized increments, the complete sampling for different spatial frequencies is obtained [34].

The MNIST database [35] was used in *Section 3.1* for training the BCNN with 800 images as the training dataset, 100 images as the validating dataset and another 100 images as the testing dataset. All the images were up-sampled from 28×28 to 32×32 to meet the dimension requirement of the RD Hadamard patterns. The full RD Hadamard basis for a 32×32 image has 1,024 RD

Hadamard patterns each with a size of 32×32. Varying compression ratios (varying levels of model ill-posedness) were used here as 8X, 16X, 32X and 64X corresponding to taking the first 1/8, 1/16, 1/32 and 1/64 of RD Hadamard patterns, respectively. The one-dimensional (1D) raw measurement data were acquired by multiplying each individual image with the RD Hadamard patterns at each compression ratio. Therefore, the 1D raw measurement data have a size of 128×1, 64×1, 32×1 and 16×1 for the corresponding compression ratios. Besides, 25dB signal-to-noise ratio (SNR) of white Gaussian noise was added to the 1D measurement data.

The STL-10 natural image database [36] was used in *Section 3.3* for training the BCNN with 10,000 images as the training dataset, 2,000 images as the validating dataset and another 2,000 images as the testing dataset. All the images were down-sampled from 96×96 to 64×64 to meet the dimension requirement of the RD Hadamard patterns. The full RD Hadamard basis for a 64×64 image has 4,096 RD Hadamard patterns each with a size of 64×64. Varying compression ratios (varying levels of model ill-posedness) were used here as 2X, 4X, 8X and 16X. Therefore, the 1D raw measurement data have a size of 2,048×1, 1,024×1, 512×1 and 256×1 for the corresponding compression ratios. Besides, 25dB SNR of white Gaussian noise was added to the 1D measurement data.

In the pre-processing step, an initial guess of each image in the training, validating and testing datasets is reconstructed using an iterative $L_2$ norm minimization approach LSQR [37], and then used as the input of BCNN for further training and prediction.

For the BCNN trained with the MNIST database, the Adam optimizer was used with a linearly decreasing learning rate starting from $5 \times 10^{-3}$ and ending with $5 \times 10^{-5}$. The batch size was chosen to be 40 and the BCNN was trained for 500 epochs. For the BCNN trained with the STL-10 database, the Adam optimizer was used with a constant learning rate of $5 \times 10^{-3}$. The batch size was chosen to be 50 and the BCNN was trained for 200 epochs.

## 3. Analysis of BCNN to approximate uncertainty in simulated SPI

In this section, we analyze BCNN in simulated SPI for uncertainty approximation. The BCNN predictions with Bernoulli-distributed, Laplacian-distributed and Gaussian-distributed likelihood functions in the simulated SPI (with varying compression ratios) trained with the MNIST database is analyzed in *Section 3.1*. The effect of noise in BCNN predictions is analyzed in *Section 3.2*. A more challenging task, the BCNN's training and predictions using the STL-10 database with Laplacian-distributed, Gaussian-distributed and Bernoulli-distributed likelihood functions is analyzed in *Section 3.3*.

## 3.1. The BCNN predictions in the simulated SPI trained with the MNIST database

In this section, we explore BCNN predictions in the simulated SPI trained with the MNIST database with Bernoulli-distributed, Laplacian-distributed and Gaussian-distributed likelihood functions. The data simulation and pre-processing are shown in *Section 2.3*. Figure 2(a) show a representative ground-truth image in the testing dataset, input images to the network from LSQR approach and BCNN predictions (with all three likelihood functions) at 8X, 16X, 32X and 64X compression ratios. To quantitatively compare the BCNN predictions with the three likelihood functions, the mean and standard deviation of the MAE and SSIM for all the predicted images, and the correlation coefficient ($R^2$) between the true absolute error and the predicted uncertainty in the testing dataset at each compression ratio were calculated and shown in Fig. 2(b-d).

Both the qualitative and quantitative results show that the accuracy of the predicted images from the three likelihood functions decreases as the compression ratio increases This is verified with an increase of the true absolute error (difference between the ground-truth image and the predicted image) in Fig. 2(a), an increase of the MAE in Fig. 2(b) and a decrease of the SSIM in Fig. 2(c). This is reasonable since higher compression ratio means higher model ill-posedness which results in solving a more difficult imaging inverse problem. However, the predicted images in BCNN with the Bernoulli-distributed likelihood function are more accurate than those with the Laplacian-distributed and Gaussian-distributed likelihood functions, especially at higher compression ratios. In terms of the predicted uncertainties, the BCNN with the Bernoulli-distributed likelihood function still performs better than the ones with Laplacian-distributed and Gaussian-distributed likelihood functions. In the BCNN with the Bernoulli-distributed likelihood function, the predicted uncertainties generally match well with the true absolute error. The regions of the predicted image from BCNN with larger errors are generally marked with higher uncertainty values in the predicted uncertainty. It can be observed from the true absolute error and predicted uncertainty that most of the higher inaccuracies come from the edges of the image features. However, the predicted uncertainties in the BCNN with Laplacian-distributed and Gaussian-distributed likelihood functions do not match well with the true absolute error. For instance, as shown in Fig. 2(a) at the 8X compression ratio, higher true absolute errors in the predicted images of BCNN with the two likelihood functions mostly come from the edges of the image features while the predicted uncertainties indicate higher uncertainties not only on the edges but also within the feature regions. The outperformance of the uncertainty predictions with the Bernoulli-distributed likelihood function can also be quantitatively seen in Fig. 2(d) with a generally higher $R^2$ value. In summary, for the MNIST database, the BCNN with the Bernoulli-distributed likelihood function performs the best among the BCNNs with three distribution likelihood functions. The BCNNs with the Laplacian-distributed and Gaussian-distributed likelihood functions are not suitable for SPI with the MNIST database. This is reasonable since the images in the MNIST database are binary which fits with the Bernoulli distribution.

In addition, it is observed that the data uncertainty is dominant over the model uncertainty. This effect becomes more pronounced at higher compression ratios. This can be shown quantitatively by the averaged pixel values in the predicted data and model uncertainties in the testing dataset in Fig. 2(e). We hypothesize that this comes from the compressed nature of the measurement data in the training set of the MNIST database.

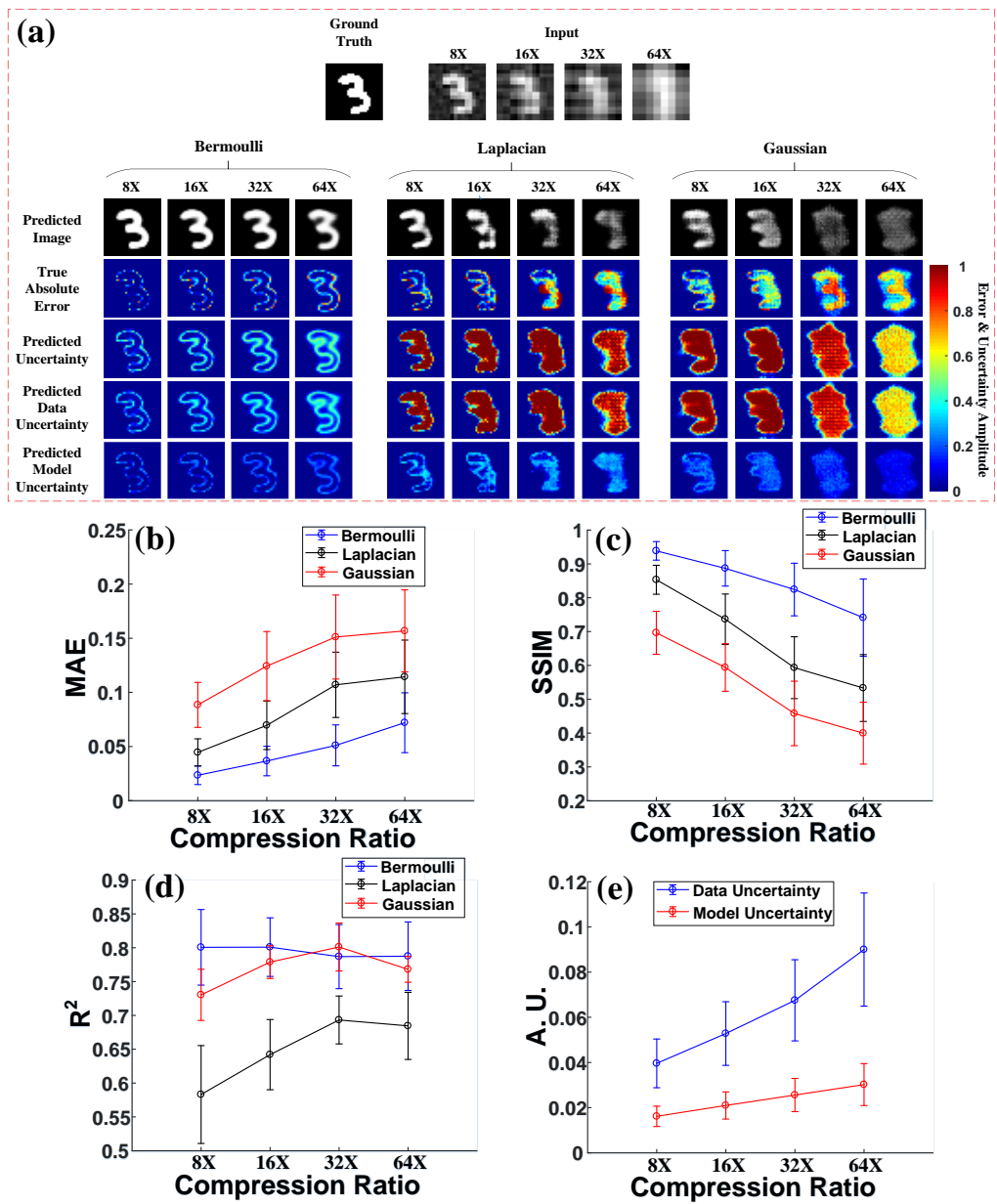

Fig. 2 The BCNN predictions in the simulated SPI trained with the MNIST database. (a) A representative ground-truth image in the testing dataset, input images to the BCNN from LSQR calculation and the BCNN predictions with Bernoulli-distributed, Laplacian-distributed and Gaussian-distributed likelihood functions at the 8X, 16X, 32X and 64X compression ratios; (b) The MAEs of the predicted images in BCNN with the three likelihood functions at the four compression ratios; (c) The SSIMs of the predicted images in BCNN with the three likelihood functions at the four compression ratios; (d) The $R^2$ of the predicted images in BCNN with the three likelihood functions at the four compression ratios; (e) Averaged pixel values of the predicted data and model uncertainties in the testing dataset at the four compression ratios.

## 3.2. Effects of noise to the BCNN performance

Since noise is one of the major challenges in imaging inverse problems, we sought to explore the effects of noise to both the image and uncertainty predictions in BCNN. The data simulation is the same as that with the MNIST database in *Section 2.3* except that varying SNR levels (0dB, 5dB, 10dB, 15dB, 20dB and 25dB) of white Gaussian noise were added to the 1D measurement data at the 16X compression ratio. The BCNN predictions are shown in Fig. 3(a). The input images from the LSQR approach suffer at higher SNR level of noise (added to the 1D measurement data) while the predicted images from BCNN is more robust although with a higher true absolute error. Besides, the predicted uncertainty generally matches well with the true absolute error at varying SNR levels. The regions where the predicted image from BCNN has larger errors are generally marked with higher uncertainty values in the predicted uncertainty. Similar to the findings in *Section 3.1*, it can still be observed from the true absolute error and predicted uncertainty that most of the higher inaccuracies come from the edges of the image features.

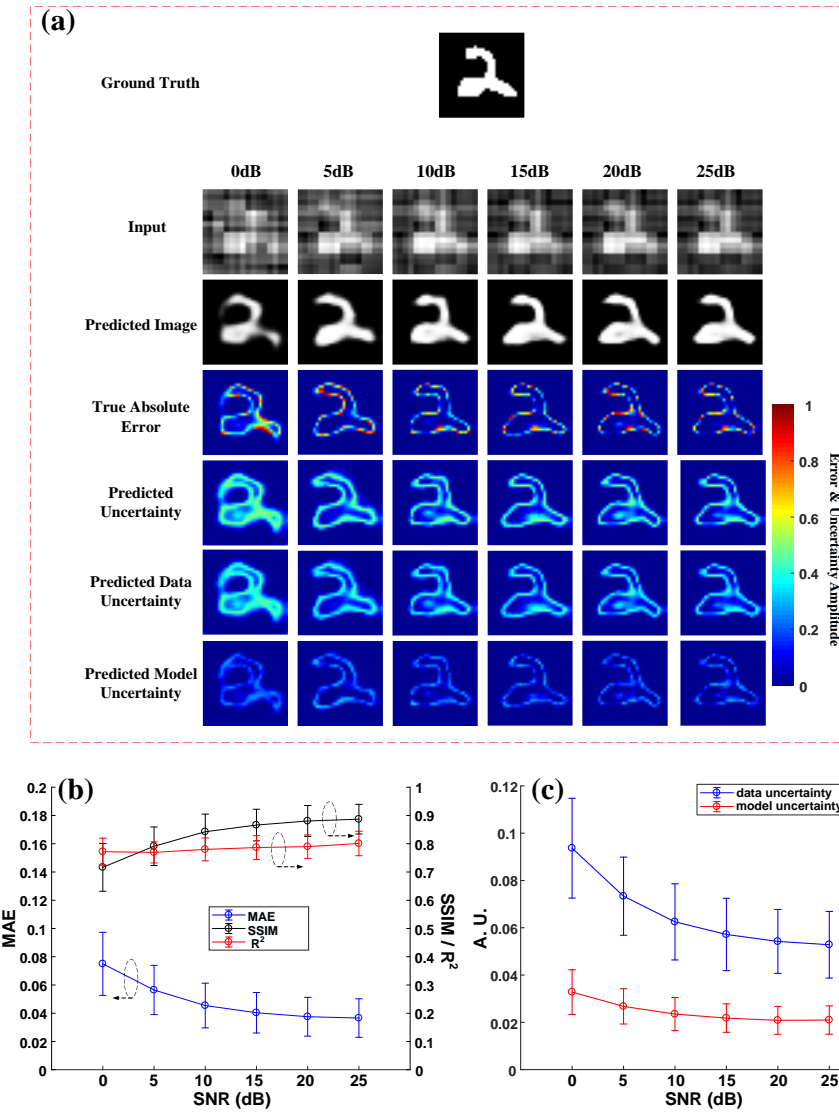

Fig. 3 Effects of noise to the BCNN performance. (a) The BCNN results for a representative image in the testing dataset with varying SNR of noise added to the raw measurement data; (b) MAE, SSIM and $R^2$ of the BCNN results with varying SNR of noise added to the raw measurement data; (c) Averaged pixel values of the predicted data and model uncertainties in the testing data with varying SNR.

To quantitatively compare the performance, the mean and standard deviation of the MAE and SSIM for all the predicted images, and the $R^2$ between the true absolute error and the predicted uncertainty in the testing dataset at each noise level were calculated as shown in Fig. 3(b). It shows that the performance of BCNN decreases as the SNR decreases from 25dB to 0dB. However, it has a good level of robustness to noise with an MAE lower than 0.08, an SSIM higher than 0.7 and a $R^2$ higher than 0.75 at the 0dB SNR level. In addition, it is observed that the data uncertainty is dominant over the model uncertainty at higher SNR levels of noise, which can be shown quantitatively in Fig. 3(c). This is reasonable since higher noise levels lead to higher variabilities in the data.

## 3.3. The BCNN predictions in the simulated SPI trained with the STL-10 database

In this section, we explore the BCNN performances with the three likelihood functions in the simulated SPI with a more challenging task where the STL-10 database with more complexed image features is used for training and predictions. The data simulation and pre-processing are shown in *Section 2.3*. Figure 4(a) show a representative ground-truth image in the testing dataset, input images to the network from the LSQR approach and Fig. 4(b) shows BCNN predictions (with all three likelihood functions) at 2X, 4X, 8X and 16X compression ratios. The mean and standard deviation of the MAE and SSIM for all the predicted images, and the $R^2$ between the true absolute error and the predicted uncertainty in the testing dataset with the three likelihood functions at each compression ratio were calculated and shown in Table 1.

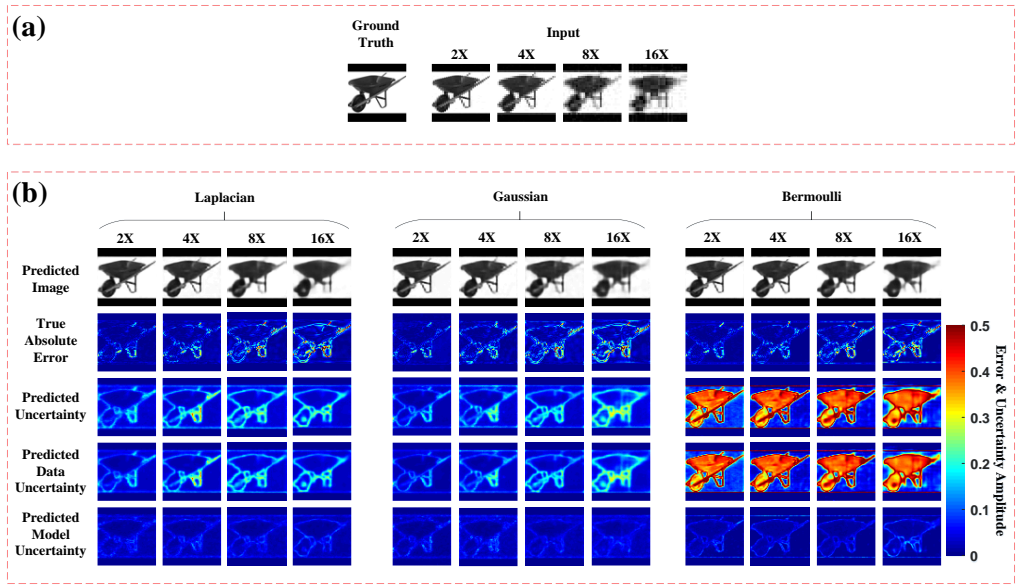

Fig. 4 The results of BCNN with Laplacian-distributed, Gaussian-distributed and Bernoulli-distributed likelihood functions in simulated SPI with STL-10 dataset at 2X, 4X, 8X and 16X compression ratios. (a) A representative ground-truth image in the testing dataset, input images to the BCNN from LSQR calculation; (b) BCNN predictions with Laplacian-distributed, Gaussian-distributed and Bernoulli-distributed likelihood functions at the 2X, 4X, 8X and 16X compression ratios.

The results in Fig. 4 and Table 1 show that the accuracy of the predicted images from the three likelihood functions decreases as the compression ratio increases, which is reasonable since higher compression ratio means higher model ill-posedness which results in solving a more difficult imaging inverse problem. Besides, the prediction of the images in BCNNs with the three likelihood functions performs close to each other as shown qualitatively in the predicted-image rows in Fig. 4(b) and quantitatively in terms of MAE and SSIM in Table 1. The predicted uncertainties in BCNNs with the Laplacian-distributed and Gaussian-distributed likelihood functions match well with the true absolute error since the regions where the predicted image from BCNN has larger errors are generally marked with higher uncertainty values in the predicted uncertainty. However, the predicted uncertainties in BCNN with the Bernoulli-distributed likelihood function are much worse as shown qualitatively in Fig. 4(b) where the low true-absolute-error pixels are marked with higher uncertainty values in the predicted uncertainty instead, and quantitatively in Table 1 where the $R^2$ values between the true absolute error and the predicted uncertainty from the BCNN with the Bernoulli-distributed likelihood function are much lower than those with the Laplacian-distributed and Gaussian-distributed likelihood function. This is reasonable since the images in the STL-10 database are in gray scale which fits better with the Laplacian and Gaussian distributions. It also indicates that the Bernoulli-distributed likelihood function is not suitable to use in BCNN when making predictions in SPI with the STL-10 database. Similar to the observations in *Section 2.1*, it can still be observed from the true absolute error and predicted uncertainty that most of the higher inaccuracies come from the edges of the image features, and that the data uncertainty is dominant over the model uncertainty.

**Table 1. Quantitative comparisons among BCNN predictions with Laplacian-distributed, Gaussian-distributed and Bernoulli-distributed likelihood functions in simulated SPI with STL-10 database.**

|     |           | MAE    | SSIM   | $R^2$  |
| --- | --------- | ------ | ------ | ------ |
| 2X  | Laplacian | 0.0218 | 0.9224 | 0.5034 |
|     | Gaussian  | 0.0212 | 0.9231 | 0.4844 |
|     | Bernoulli | 0.0205 | 0.9337 | 0.1927 |
| 4X  | Laplacian | 0.0290 | 0.8733 | 0.4778 |
|     | Gaussian  | 0.0296 | 0.8693 | 0.4865 |
|     | Bernoulli | 0.0277 | 0.8846 | 0.2240 |
| 8X  | Laplacian | 0.0383 | 0.7866 | 0.4689 |
|     | Gaussian  | 0.0385 | 0.7882 | 0.4782 |
|     | Bernoulli | 0.0367 | 0.7983 | 0.2353 |
| 16X | Laplacian | 0.0499 | 0.6661 | 0.4576 |
|     | Gaussian  | 0.0505 | 0.6680 | 0.4591 |
|     | Bernoulli | 0.0494 | 0.6761 | 0.2422 |

## 4. Discussions and conclusions

The BCNN is proposed for uncertainty approximation in SPI with Bernoulli-distributed, Laplacian-distributed or Gaussian-distributed likelihood functions with the MNIST and STL-10 databases. First, the BCNNs with the three distribution likelihood functions were compared in simulated SPI with the MNIST database at varying compression ratios and the Bernoulli-distributed likelihood function was proved to be the most appropriate among the three functions. Second, the robustness of BCNN to noise from the measurement data was studied with the Bernoulli-distributed likelihood function and MNIST database. Third, the three likelihood functions were compared in BCNN in simulated SPI with STL-10 dataset and the Laplacian-distributed and Gaussian-distributed likelihood functions were shown to be equivalent and both better than the Bernoulli-distributed likelihood function in this application.

However, several things still need further exploration. First, how to choose the optimal probability-distributed likelihood function in BCNN efficiently is a problem that needs to be explored. Currently, we train the BCNN with all the potential probability-distributed likelihood functions and then compare the results to find the optimal one. However, the drawback is that it is time consuming. We would like to propose a method to find the optimal distribution before training using the dataset statistics and network architecture. Second, it is observed that the data uncertainty is dominant over the model uncertainty in both simulations and experiments due to the compressed nature and noise in the measurement data in the training dataset. We would like to search for more advanced pre-processing approaches to decrease the uncertainty stemming from the measurement data. Third, we would like to explore ways to use the predicted uncertainty as a feedback to optimize the BCNN structures to further decrease the uncertainty of the results.

In summary, the proposed BCNN enables uncertainty approximation in SPI. It is a reliable tool in diverse applications in SPI where the confidence on the predicted images needs to be approximated.

## 5. Funding

This work was supported by NIH grant R21GM137334, a Faculty Grant from the Neukom Institute for Computational Science at Dartmouth College and the Alma Hass Milham Fellowship in Biomedical Engineering from Thayer School of Engineering at Dartmouth College.

## 6. Disclosures

The authors declare no conflicts of interest.

# References

1. G. M. Gibson, S. D. Johnson, and M. J. Padgett, "Single-pixel imaging 12 years on: a review," Optics Express **28**, 28190-28208 (2020).

2. I. Hoshi, T. Shimobaba, T. Kakue, and T. Ito, "Single-pixel imaging using a recurrent neural network combined with convolutional layers," Optics Express **28**, 34069-34078 (2020).

3. B. I. Erkmen, "Computational ghost imaging for remote sensing," JOSA A **29**, 782-789 (2012).

4. P. Clemente, V. Durán, E. Tajahuerce, P. Andrés, V. Climent, and J. Lancis, "Compressive holography with a single-pixel detector," Optics letters **38**, 2524-2527 (2013).

5. Y. Endo, T. Tahara, and R. Okamoto, "Color single-pixel digital holography with a phase-encoded reference wave," Applied optics **58**, G149-G154 (2019).

6. Z. Zhang, S. Jiao, M. Yao, X. Li, and J. Zhong, "Secured single-pixel broadcast imaging," Optics express **26**, 14578-14591 (2018).

7. S. Jiao, C. Zhou, Y. Shi, W. Zou, and X. Li, "Review on optical image hiding and watermarking techniques," Optics & Laser Technology **109**, 370-380 (2019).

8. J. Peng, M. Yao, J. Cheng, Z. Zhang, S. Li, G. Zheng, and J. Zhong, "Micro-tomography via single-pixel imaging," Optics express **26**, 31094-31105 (2018).

9. E. Candes, and J. Romberg, "Sparsity and incoherence in compressive sampling," Inverse problems **23**, 969 (2007).

10. E. J. Candès, and M. B. Wakin, "An introduction to compressive sampling [a sensing/sampling paradigm that goes against the common knowledge in data acquisition]," IEEE signal processing magazine **25**, 21-30 (2008).

11. G. Barbastathis, A. Ozcan, and G. Situ, "On the use of deep learning for computational imaging," Optica **6**, 921-943 (2019).

12. Y. LeCun, Y. Bengio, and G. Hinton, "Deep learning," nature **521**, 436 (2015).

13. Y. Rivenson, Z. Göröcs, H. Günaydin, Y. Zhang, H. Wang, and A. Ozcan, "Deep learning microscopy," Optica **4**, 1437-

1443 (2017).

14. Y. Rivenson, Y. Zhang, H. Günaydın, D. Teng, and A. Ozcan, "Phase recovery and holographic image reconstruction using deep learning in neural networks," Light: Science & Applications **7**, 17141 (2018).

15. Y. Li, Y. Xue, and L. Tian, "Deep speckle correlation: a deep learning approach toward scalable imaging through scattering media," Optica **5**, 1181-1190 (2018).

16. Y. Xue, S. Cheng, Y. Li, and L. Tian, "Reliable deep-learning-based phase imaging with uncertainty quantification," Optica **6**, 618-629 (2019).

17. A. Sinha, J. Lee, S. Li, and G. Barbastathis, "Lensless computational imaging through deep learning," Optica **4**, 1117-1125 (2017).

18. A. Goy, K. Arthur, S. Li, and G. Barbastathis, "Low photon count phase retrieval using deep learning," Physical review letters **121**, 243902 (2018).

19. A. Goy, G. Rughoobur, S. Li, K. Arthur, A. I. Akinwande, and G. Barbastathis, "High-resolution limited-angle phase tomography of dense layered objects using deep neural networks," Proceedings of the National Academy of Sciences **116**, 19848-19856 (2019).

20. C. F. Higham, R. Murray-Smith, M. J. Padgett, and M. P. Edgar, "Deep learning for real-time single-pixel video," Scientific reports **8**, 1-9 (2018).

21. M. Lyu, W. Wang, H. Wang, H. Wang, G. Li, N. Chen, and G. Situ, "Deep-learning-based ghost imaging," Scientific reports **7**, 1-6 (2017).

22. R. Shang, K. Hoffer-Hawlik, F. Wang, G. Situ, and G. P. Luke, "Two-step training deep learning framework for computational imaging without physics priors," Optics Express **29**, 15239-15254 (2021).

23. T. Shimobaba, Y. Endo, T. Nishitsuji, T. Takahashi, Y. Nagahama, S. Hasegawa, M. Sano, R. Hirayama, T. Kakue, and A. Shiraki, "Computational ghost imaging using deep learning," Optics Communications **413**, 147-151 (2018).

24. F. Wang, H. Wang, H. Wang, G. Li, and G. Situ, "Learning from simulation: An end-to-end deep-learning approach for computational ghost imaging," Optics express **27**, 25560-25572 (2019).

25. S. Jiao, J. Feng, Y. Gao, T. Lei, Z. Xie, and X. Yuan, "Optical machine learning with incoherent light and a single-pixel detector," Optics letters **44**, 5186-5189 (2019).

26. Z. Wang, A. C. Bovik, H. R. Sheikh, and E. P. Simoncelli, "Image quality assessment: from error visibility to structural similarity," IEEE transactions on image processing **13**, 600-612 (2004).

27. A. Kendall, and Y. Gal, "What uncertainties do we need in bayesian deep learning for computer vision?," arXiv preprint arXiv:1703.04977 (2017).

28. Y. Kwon, J.-H. Won, B. J. Kim, and M. C. Paik, "Uncertainty quantification using Bayesian neural networks in classification: Application to biomedical image segmentation," Computational Statistics & Data Analysis **142**, 106816 (2020).

29. Y. Gal, and Z. Ghahramani, "Dropout as a bayesian approximation: Representing model uncertainty in deep learning," in *international conference on machine learning*(PMLR2016), pp. 1050-1059.

30. B. Lakshminarayanan, A. Pritzel, and C. Blundell, "Simple and scalable predictive uncertainty estimation using deep ensembles," arXiv preprint arXiv:1612.01474 (2016).

31. N. Srivastava, G. Hinton, A. Krizhevsky, I. Sutskever, and R. Salakhutdinov, "Dropout: a simple way to prevent neural networks from overfitting," The journal of machine learning research **15**, 1929-1958 (2014).

32. B. Hanin, and D. Rolnick, "How to start training: the effect of initialization and architecture," in *Proceedings of the 32nd International Conference on Neural Information Processing Systems*(2018), pp. 569-579.

33. O. Ronneberger, P. Fischer, and T. Brox, "U-net: Convolutional networks for biomedical image segmentation," in *International Conference on Medical image computing and computer-assisted intervention*(Springer2015), pp. 234-241.

34. M.-J. Sun, L.-T. Meng, M. P. Edgar, M. J. Padgett, and N. Radwell, "A Russian Dolls ordering of the Hadamard basis for compressive single-pixel imaging," Scientific reports **7**, 3464 (2017).

35. Y. LeCun, L. Bottou, Y. Bengio, and P. Haffner, "Gradient-based learning applied to document recognition," Proceedings of the IEEE **86**, 2278-2324 (1998).

36. A. Coates, A. Ng, and H. Lee, "An analysis of single-layer networks in unsupervised feature learning," in *Proceedings of the fourteenth international conference on artificial intelligence and statistics*(2011), pp. 215-223.

37. C. C. Paige, and M. A. Saunders, "LSQR: An algorithm for sparse linear equations and sparse least squares," ACM Transactions on Mathematical Software (TOMS) **8**, 43-71 (1982).